\documentclass[conference]{IEEEtran}
\IEEEoverridecommandlockouts
\usepackage[utf8]{inputenc}
\usepackage{amsmath}
\usepackage{cite}
\usepackage{multicol}
\usepackage{graphicx}
\usepackage{array}
\usepackage{multirow}
\usepackage{graphicx}
\usepackage{enumitem}
\usepackage{color}
\usepackage{nomencl}
\usepackage{algorithm}
\usepackage{algpseudocode}
\usepackage{ifthen}
\usepackage{etoolbox}
\usepackage{amssymb}
\usepackage{comment}
\usepackage{mathrsfs}
\usepackage{mathtools}
\usepackage{bbold}
\usepackage{yfonts}
\usepackage{soul}
\usepackage{placeins}
\usepackage{fnpct}
\usepackage{courier}
\usepackage{ntheorem}
\usepackage{cuted}
\usepackage{stfloats}
\usepackage{hyperref}

\usepackage{fnpct}
\ifCLASSOPTIONcompsoc \usepackage[caption=false,font=normalsize,labelfon
t=sf,textfont=sf]{subfig}
\else
\usepackage[caption=false,font=footnotesize]{subfig}
\fi
\usepackage{tikz}

\def\BibTeX{{\rm B\kern-.05em{\sc i\kern-.025em b}\kern-.08em
    T\kern-.1667em\lower.7ex\hbox{E}\kern-.125emX}}

\title{Measurement-based/Model-less Estimation of Voltage Sensitivity Coefficients by Feedforward and LSTM Neural Networks in Power Distribution Grids}

\author{
\IEEEauthorblockN{Robin Henry$^{\dagger}$, Rahul Gupta$^{\dagger, *}$}
{$^{\dagger}$Swiss Federal Institute of Technology, Lausanne Switzerland}\\
{$^{*}$School of Electrical and Computer Engineering, Georgia Institute of Technology,} Atlanta, USA \\
robin@robinxhenry.com, rgupta460@gatech.edu
}

\usepackage{etoolbox}
\makeatletter
\patchcmd{\@maketitle}
  {\addvspace{0.5\baselineskip}\egroup}
  {\addvspace{-0.5\baselineskip}\egroup}
  {}
  {}
\makeatother
\begin{document}
\maketitle

\begin{abstract}
The increasing adoption of measurement units in electrical power distribution grids has enabled the deployment of data-driven and measurement-based control schemes. Such schemes rely on measurement-based estimated models, where the models are first estimated using raw measurements and then used in the control problem.
This work focuses on measurement-based estimation of the voltage sensitivity coefficients which can be used for voltage control. In the existing literature, these coefficients are estimated using regression-based methods, which do not perform well in the case of high measurement noise. This work proposes tackling this problem by using neural network (NN)-based estimation of the voltage sensitivity coefficients which is robust against measurement noise. In particular, we propose using Feedforward and Long-Short Term Memory (LSTM) neural networks. The trained NNs take measurements of nodal voltage magnitudes and active and reactive powers and output the vector of voltage magnitude sensitivity coefficients.
The performance of the proposed scheme is compared against the regression-based method for a CIGRE benchmark network.
\end{abstract}


\vspace{-0.5em}
\section{Introduction}
Distribution system operators (DSOs) are progressively integrating measurement and monitoring infrastructures into their networks to facilitate automated control of flexible resources for executing several control functionalities such as voltage control \cite{agalgaonkar2013distribution} and congestion management \cite{christakou2017voltage}. Such control schemes suffer from two main challenges. The \emph{first} involves addressing the high computational burden of grid-aware control schemes, also referred to as the optimal power flow (OPF) problem. The \emph{second} challenge pertains to the imperative need for accurate knowledge of grid models, which may not be readily available or, if accessible, may not be accurate.

Recent literature has addressed the above two problems by developing model-less or measurement-based control schemes using the First-order Taylor's approximation of the original AC power flow equations, as exemplified in \cite{carpita2019low, su2019augmented, zhang2017locally, gupta2022model, gupta2023experimental}. These works rely on measurement-based estimation schemes; however, the regression-based method used in these works is very sensitive to the measurement noise of the voltage and current sensors, as pointed out in \cite{zhang2017locally, gupta2022model}.
This paper proposes to use neural networks (NN) for estimating the sensitivity coefficients which are robust against measurement noise. Specifically, we train NNs that can output the vectors of voltage sensitivity coefficients (with respect to the active and reactive powers) corresponding to the measurements of nodal voltage magnitudes, active and reactive powers. 

In the existing literature, the estimation of voltage sensitivity coefficients has been well-researched. The works in \cite{mugnier2016model, da2019data} used least-squares-based estimation approaches, in \cite{su2019augmented, gupta2022model} recursive least-squares methods are proposed. The work in \cite{zhang2017locally} proposed ridge-regression-based methods and weighted least squares method. However, all these schemes tend to perform worse with the increase in the measurement noise. Additionally, these schemes do not perform well in the case of multicollinear measurements (i.e., when the measured powers at different nodes show similar variations). 

Recently, NN-based approaches have been widely used for training miscellaneous control policies. For example, the work in \cite{pourjafari2019support} trained NNs to develop a voltage control model for distribution grids. In \cite{hu2020artificial, li2022day}, a model for dispatching was proposed. In \cite{li2018artificial}, volt-var control NNs were proposed. However, all these schemes are not generic enough to be applied to any distribution network (or any control policy) and require to be retrained. Additionally, all these schemes rely on accurate knowledge of the grid parameters and topology, which may not be available in real life.

In this respect, this work\footnote{The work was originally presented as part of the author's thesis work \cite{Robin_thesis}.} proposes NN-based approaches for the measurement-based estimation of the voltage sensitivity coefficient without knowledge of the network's parameters. NNs are trained on noisy measurements of nodal voltage magnitudes, and active and reactive powers to estimate the voltage sensitivity coefficients. The estimation problem is not straightforward as we don't have access to the true sensitivity coefficients, assuming that we do not know the grid parameters. To address this challenge, we propose using a loss function that compares the voltage estimation with respect to the measured values. We apply two widely used NNs: Feedforward Neural Network (FNN) and Long-Short Term Memory (LSTM). The method is numerically validated on the CIGRE microgrid benchmark system connected with multiple photovoltaic (PV) plants. We compare the performances of the proposed NN-based methods to conventional regression-based methods. We also compare the performance for different levels of measurement noise, as per the characteristics of the instrument transformer (IT) classes.

The paper is organised as follows. Section~\ref{sec:prob_stat} presents an overview of the sensitivity estimation problem, Sec.~\ref{sec:NN} presents the NN-based sensitivity estimation schemes, Sec.~\ref{sec:sim_res} presents the setup for simulation and the numerical results, and finally Sec.~\ref{sec:conclusion} presents the conclusion.

\section{Sensitivity Coefficient Estimation Problem}
\label{sec:prob_stat}
We consider a power distribution network for which the network model, i.e., the line parameters, is not known. We assume that the network is equipped with measurement instruments such as smart meters (SMs) or phasor measurement units (PMUs), providing voltage magnitudes and power measurements, and the locations of the SMs/PMUs are known. The objective is to estimate the sensitivity coefficients of voltage magnitudes using recent voltage and power measurements. 

Let the distribution network consist of $n_b$ nodes. Let the symbol $V_i$ denote the voltage phasor for $i$ node, and the complex vector $\mathbf{V} \in \mathbb{R}^{n_b-1}$ contains the voltage phasors of all nodes (except the slack node). The vector $|\mathbf{V}|$ represents the voltage magnitudes. Additionally, let the symbols $\mathbf{P} = [P_1, \dots, P_{n_b-1}] \in \mathbb{R}^{n_b-1}$ and $\mathbf{Q} = [Q_1, \dots, Q_{n_b-1}] \in \mathbb{R}^{n_b-1}$ denote the active and reactive powers, respectively.
Using, the power flow equations, each nodal voltage can be expressed as non-linear function $\mathcal{V}_i: \mathbb{R}^{2n_b-1} \rightarrow \mathbb{R}$ of power injections and the slack voltage ($V_0$) as
\begin{align}
    |V_i| = \mathcal{V}_i(V_0, \mathbf{P}, \mathbf{Q}) \label{eq:V_nonlinear}
\end{align}
Using the First order Taylor's approximation, \eqref{eq:V_nonlinear} is often linearized as 
\begin{align}
    |\Delta V_i| \approx \sum_{j=1}^{n_b-1}\frac{\partial{|V_i|}}{\partial {P_j}}\Delta {P_j} + \sum_{j=1}^{n_b-1}\frac{\partial{|V_i|}}{\partial {Q_j}} \Delta{Q_j} \label{eq:lin_vmodel}
\end{align}
where, $K^p_{ij} = \frac{\partial{|V_i|}}{\partial {P_j}}$ and $K^q_{ij} = \frac{\partial{|V_i|}}{\partial {Q_j}}$ are the voltage magnitude sensitivity coefficients. The symbols $\Delta P_j$ and $\Delta Q_j$ denote deviation of the nodal active and reactive powers at node $j$.

In a measurement-based estimation setting, the voltage and power magnitudes are subjected to the measurement noise as per the characteristics of the instrument transformers (ITs) \cite{IT_C, IT_V} and are represented by $\tilde{V}, \tilde{P}, \tilde{Q}$, modeled as 
\begin{align}
    \tilde{V} = V + \Delta{V},~ \tilde{P} = P + \Delta{P},~ \tilde{Q} = Q + \Delta{Q}, 
\end{align}
where $\Delta(.)$ refers to the measurement sensor noise.

Assuming that the measurements of the nodal voltages are available for several time-steps indexed by $m = 1, \dots, M$, and the coefficients do not change during this period, \eqref{eq:lin_vmodel} can be expressed for $M$ timesteps as

\small
\begin{align}
\underbrace{
\begin{bmatrix}
    |\Delta \tilde{V}_{i,1}|\\
    \vdots \\
    |\Delta \tilde{V}_{i,m}| \\
    \vdots \\
    |\Delta \tilde{V}_{i,M}|
\end{bmatrix}}_{\mathbf{y}_i} \approx
\underbrace{
    \begin{bmatrix}
        \Delta \tilde{\mathbf{P}}_1 & \Delta \tilde{\mathbf{Q}}_1 \\
         \vdots & \vdots \\
        \Delta \tilde{\mathbf{P}}_m & \Delta \tilde{\mathbf{Q}}_m \\
         \vdots  & \vdots \\
        \Delta \tilde{\mathbf{P}}_M &  \Delta \tilde{\mathbf{Q}}_M\\
    \end{bmatrix}}_{\mathbf{A}_i}
    \underbrace{
    \begin{bmatrix}
        K^p_{i1}\\ \vdots \\ K^p_{i(n_b-1)}\\ K^q_{i1} \\ \dots \\ K^q_{i(n_b-1)}
    \end{bmatrix}}_{\boldsymbol{z}_i}
    \label{eq:measur_model}
\end{align}

\normalsize
In \eqref{eq:measur_model}, $\mathbf{y}_i$ and $\mathbf{A}_i$ are known from the measurements, whereas $\mathbf{z}_i$ is the variable to be estimated. This problem is often solved by least squares (LS) regression as in \cite{mugnier2016model, zhang2017locally} and given by 
\begin{align}
    \widehat{\mathbf{z}}_i = (\mathbf{A}_i^\top \mathbf{A}_i)^{-1}\mathbf{A}_i^\top \mathbf{y}_i
\end{align}
As reported in previous works (e.g., \cite{zhang2017locally, gupta2022model}), LS doesn't perform well with an increase in measurement noise. To address this issue, in the following, we present two neural network approaches for estimating sensitivity coefficients that are resistant to measurement noise.
\section{Neural Network-based Estimation}
\label{sec:NN}
\begin{figure}[!htbp]
    \centering
    \includegraphics[width=0.95\linewidth]{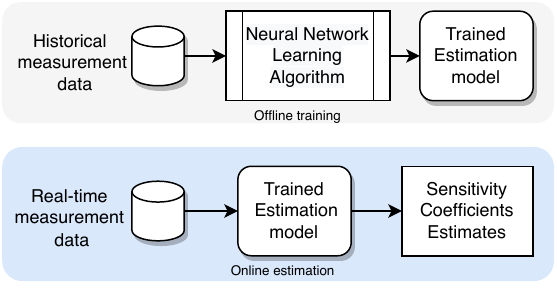}
    \caption{Schematic diagram for the proposed NN-based training and estimation scheme for sensitivity coefficients.}
    \label{fig:flowchart}
\end{figure}

We propose solving the estimation problem through NN-based frameworks that can handle measurement noise and multicollinear datasets. The scheme is depicted in Fig.~\ref{fig:flowchart}, where a training dataset is initially employed to train the NN. Subsequently, the trained NN is utilized for estimation on unseen real-time measurements.

We formulate the sensitivity coefficient estimation problem using two different supervised NN algorithms, essentially mapping voltage and power measurements to voltage magnitude sensitivity coefficients. However, as the grid model (line parameters) is unknown, we lack access to the \emph{true} sensitivity coefficients before estimation. To address this challenge, we designed a surrogate loss function $\mathcal{L}$ that evaluates the accuracy of voltage estimation. Consequently, it only relies on the available measurements and does not necessitate the knowledge of true sensitivity coefficients during training. The loss function is expressed as
\begin{align}
\label{eq:loss_func}
    & \mathcal{L}(\mathbf{y}_i, \hat{\mathbf{y}}_i) = \|\mathbf{y}_i - \hat{\mathbf{y}}_i\|_2^2, ~ \text{where},\\
    & 
    \hat{\mathbf{y}} = 
    \mathbf{A}_i\hat{\mathbf{z}}_i
\end{align}
where $\hat{\mathbf{z}}_i$ is the vector of estimated coefficients by NN.

Applying the aforementioned loss function, we utilize two NN approaches: the first is an FNN, and the second is an LSTM network. They are described as follows.
\subsection{Feedforward Neural Network (FNN)}
FNN is one of the most widely used supervised learning algorithms which maps a set of pre-defined inputs to a particular output. 
\begin{figure}[!htbp]
    \centering
    \includegraphics[width=0.95\linewidth]{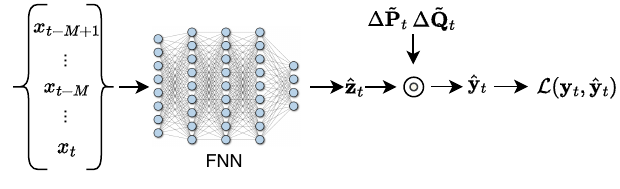}
    \caption{{FNN-based training model of the sensitivity coefficient estimation.}}
    \label{fig:neural_network}
\end{figure}
FNN consists of a series of layers;  
each layer takes inputs from the previous layer, applies a weighted linear function, and passes the result to the next layer.

For the sensitivity coefficient estimation problem, separate FNNs are trained for each non-slack node. In each FNN, the input comprises measurements from a time window of length $M$. Each timestep includes nodal voltage magnitudes, and active and reactive power measurements, represented by a vector $\mathbf{x} = [x_{t-M+1}, \dots, x_{t}]^{\top} \in \mathbb{R}^{2(n_b-1)M + 1}$ where $x_t = [\Delta{V}_{i,t}, \mathbf{P}_t, \mathbf{Q}_t]^{\top}$. The FNN output corresponds to the sensitivity coefficients $\hat{\mathbf{z}}_i \in \mathbb{R}^{2(n_b-1)}$, providing estimates of the voltage sensitivity coefficients for a specific node. To estimate coefficients for all non-slack nodes, separate parallel FNNs are trained. Fig.~\ref{fig:neural_network} illustrates a schematic of the FNN training process.
In our case, we considered FNNs with two fixed hidden layers of sizes 128 and 64 neurons, respectively. All layers, except the output one, were followed by a ReLU activation function.
\subsection{Long-Short Term Memory (LSTM) Neural Network} The existing literature, as extensively reported (e.g., \cite{brezak2012comparison}), suggests that FNNs may perform poorly for time series data, such as those for voltage and power measurements. An alternative approach is to use Recurrent Neural Networks (RNNs), which have demonstrated better performance for time series data. RNNs consist of an internal hidden state that gets updated after each estimation, making them ideal for tasks that require contextual information or a memory component.

\begin{figure}[!htbp]
    \centering
    \includegraphics[width=0.9\linewidth]{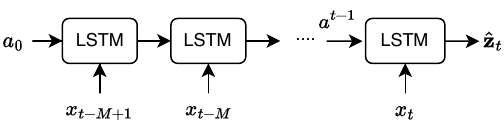}
    \caption{LSTM-based training model of the sensitivity coefficient estimation.}
    \label{fig:lstm}
\end{figure}

An example of an RNN-based network is LSTM \cite{greff2016lstm}, which we use for the sensitivity estimation problem. Fig.~\ref{fig:lstm} depicts a schematic for the LSTM approach, where information is passed to the next LSTM layer along with new measurements.
As shown in Fig.~\ref{fig:lstm}, in contrast to FNN where a window of size $M$ measurements are fed at once, in the case of LSTM, the measurement data is fed successively for each timestep and passed to the next LSTM cell.
\subsection{Training and validation procedure} 
The objective of training the NN is to ensure its performance on previously unseen data points, i.e., that it generalizes well. 
Real measurements from a given distribution network (detailed in Sec. \ref{sec:sim_res}) are used for the training and validation of NN-based algorithms, with the data divided into training and validation sets. 
The network is trained on the training set, and its performance is regularly evaluated on the held-out validation set. 
The training procedure is often adapted based on the resulting validation accuracy to minimize overfitting and ensure good generalization performance \cite{hawkins2004problem}. %
The \emph{training} process is identical for both FNN and LSTM networks, and is described as follows:
\begin{enumerate}[label=(\roman*)]
    \item Sample a pair $(\mathbf{x}_i, \mathbf{y}_i)$ from the dataset and pass it through the NNs to obtain the output $\hat{\mathbf{y}}_i.$
    \item Compute the loss function $\mathcal{L}(\mathbf{y}, \hat{\mathbf{y}})$ as per \eqref{eq:loss_func}.
    \item Update the NN parameters based on the computed loss.
    \item Repeat steps 2-4 until convergence.
\end{enumerate}

The training process outlined above is usually conducted in terms of batches and epochs. A \emph{batch} of training samples, denoted as $\mathcal{B}$, is a uniformly sampled subset of the training dataset $\mathcal{D}$ that is fed into the NN (steps 1-2). Subsequently, the parameters of the corresponding NN are updated in step 3. Updating the parameters every $|\mathcal{B}|$ steps has been shown to significantly reduce training time and promote better convergence \cite{goodfellow2016deep}. An \emph{epoch} refers to a complete pass through the dataset. In other words, if an FNN is trained for 10 epochs using a dataset of $\mathcal{D}$ input-output pairs and a batch size of $\mathcal{B}$, the neural network parameters will be updated $10 \times \mathcal{D}/\mathcal{B}$ times. The NN parameters are updated through the backpropagation algorithm, which is a gradient-based optimization scheme \cite{rojas1996backpropagation}.

After each epoch of training, the \emph{validation} loss was computed as the average loss on the validation set. Whenever a new lowest validation loss was encountered, the parameters of the neural network were saved. After 20 epochs, the best model was selected and retained. 
During training, the Adam optimizer \cite{kingma2014adam} was used with a learning rate of 1e-3. 

\section{Simulation results}
\label{sec:sim_res}
The following presents the numerical validation of the proposed framework. First, we describe the simulation setup, providing details about the grid and the measurement noise model. Subsequently, we present the estimation results of the voltage sensitivity coefficients and compare the performance against the least squares (LS) method. Additionally, we analyze the estimation performance sensitivity to increasing measurement noise. The code can be accessed at out Github \emph{repository} \cite{Robin_thesis}.
\subsection{Simulation setup}
\subsubsection{Grid data}
\begin{figure}[!htbp]
    \centering
    \includegraphics[width=0.95\linewidth]{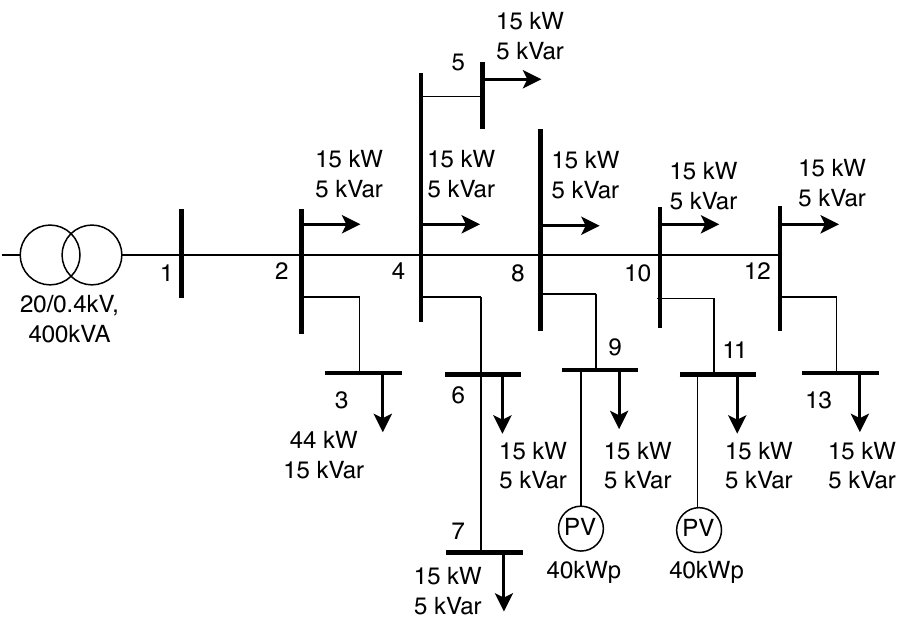}
    \caption{CIGRE Microgrid benchmark network.}
    \label{fig:cirge_grid}
\end{figure}

We validate the proposed schemes on a network for which the network parameters are fully known, enabling us to validate the estimated coefficients against the \emph{true} ones (computed analytically using methods described in \cite{christakou2013efficient}). The proposed estimation scheme is validated on the CIGRE microgrid benchmark network \cite{CIGREREF}, a low voltage 400V, 400 kV three-phase balanced distribution network. The grid layout is shown in Fig.~\ref{fig:cirge_grid}, illustrating the nominal power demands and photovoltaic (PV) generation.

\begin{figure}[!htbp]
    \centering
    \subfloat[Active power load at different nodes.]{\includegraphics[width=\columnwidth]{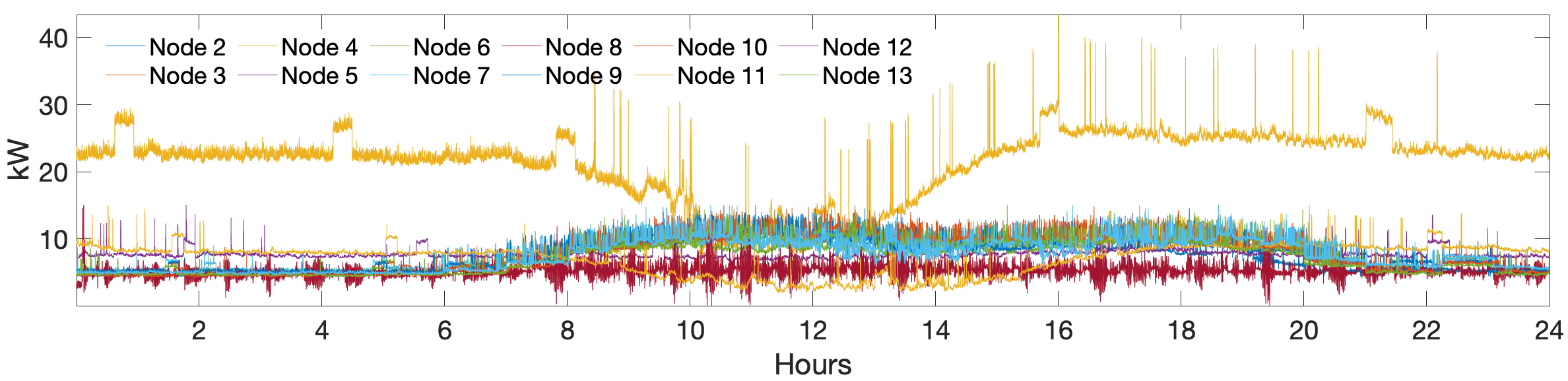}
    \label{fig:Pnodal}}\\
    \subfloat[Reactive power load at different nodes.]{\includegraphics[width=\columnwidth]{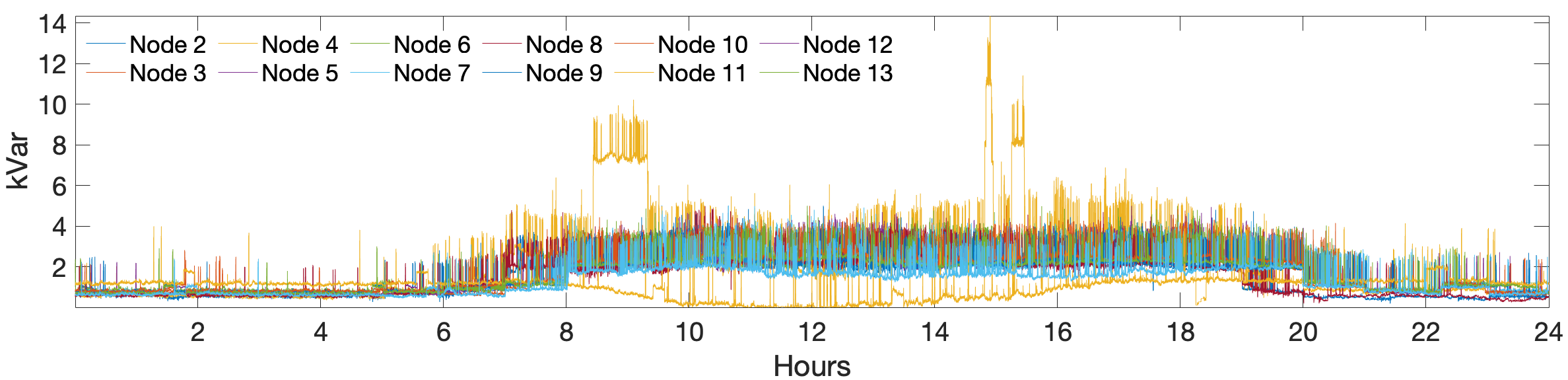}
    \label{fig:Qnodal}}\\
    \subfloat[PV generation at different nodes.]{\includegraphics[width=\columnwidth]{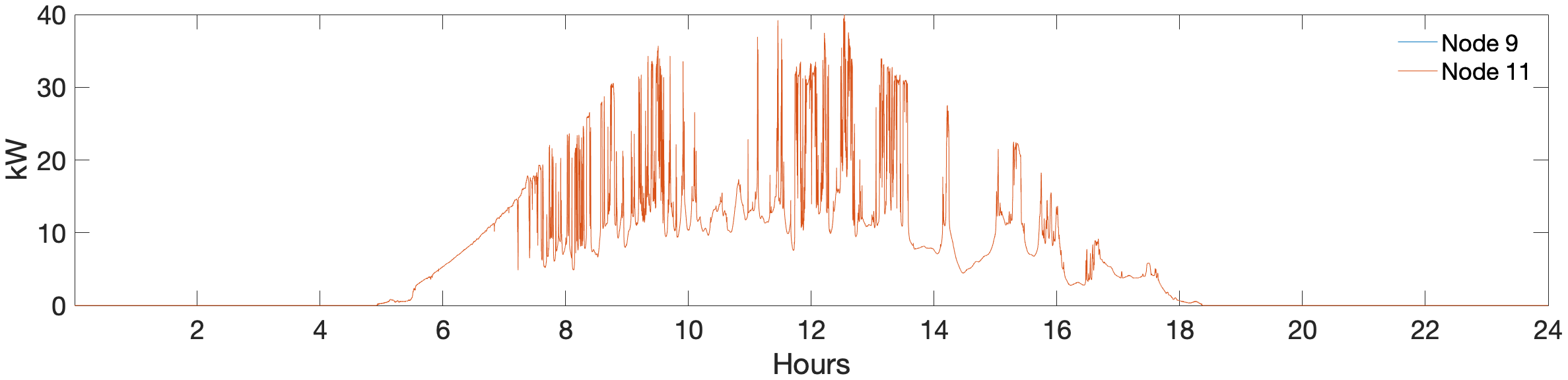}
    \label{fig:Pgen}}\\
    \subfloat[Voltage magnitudes at different nodes.]{\includegraphics[width=\columnwidth]{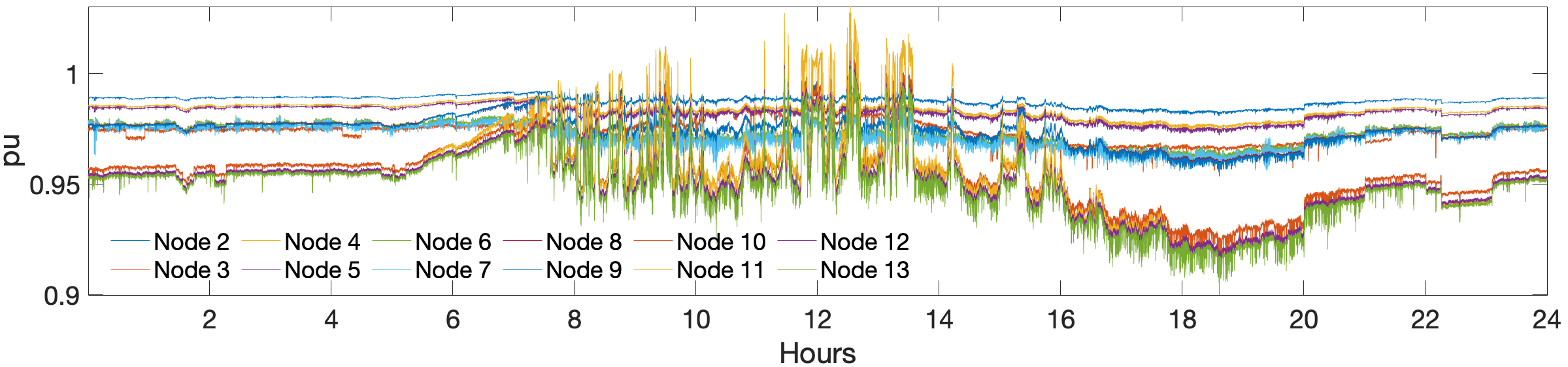}
    \label{fig:Vnodal}}
    \caption{Daily demand profiles, PV generation, and voltage magnitudes at different nodes, used for the numerical validation of the proposed estimation algorithm.}
    \label{fig:PF_profiles}
\end{figure}

The network is simulated with real active and reactive power profiles measured at the microgrid setup at the Distributed Electrical Systems Laboratory (DESL), EPFL Switzerland, using the setup described in \cite{gupta2020grid, gupta2023experimental}. The active and reactive demand profiles are illustrated in Fig.\ref{fig:Pnodal} and \ref{fig:Qnodal}, respectively. Additionally, we simulate PV generation using real measurements from the same experimental setup, and the profiles are shown in Fig.\ref{fig:Pgen}. It is worth noting that the grid hosts two PV plants, leading to the challenge of multi-collinearity, as described in \cite{zhang2017locally, gupta2022model}, due to the similarity in the variation of power injection at these nodes.
Fig.~\ref{fig:Vnodal} displays the voltage magnitudes at different nodes without measurement noise. All the profiles are available with a sampling rate of 1 second
\subsubsection{Measurement noise model}
The power and voltage profiles are subjected to realistic measurement noise, which is simulated by introducing noise in polar coordinates (i.e., magnitudes and phase) for the voltage and currents, as specified by the instrument transformers (ITs). These specifications are defined in \cite{IT_C, IT_V} for different IT classes, as outlined in Table~\ref{tab:instrument_class}. The steps for introducing noise to the magnitudes and phase measurements are detailed in Algorithm~\ref{alg:GenData}. For each time step, load flow is conducted to compute the voltage and currents, followed by the introduction of noise in polar coordinates for a given IT class. Subsequently, active and reactive powers are computed with the noisy voltage and current phasors.

\begin{table}[!htbp]
\caption{Errors specifications for different Instrument Transformers classes defined by \cite{IT_V, IT_C}.}
\begin{center}
\begin{tabular}{ |c|c|c|c|c|} 
\hline
\textbf{IT class} & \multicolumn{2}{|c|}{\textbf{Voltage transformers}} & \multicolumn{2}{|c|}{\textbf{Current transformers}}  \\
\hline
\textbf{} & \textbf{mag. error} & \textbf{phase error} & \textbf{mag. error} & \textbf{phase error}  \\
 & (${\sigma^m}$) {[\%]} &  $(\sigma^p)$ {[rad.]} &  (${\sigma^m}$) {[\%]} &  $(\sigma^p)$ {[rad.]} \\
  \hline
         0.2 & 0.2 & 3e-3 & 0.2 & 3e-3 \\
         0.5 & 0.5 & 6e-3 & 0.5 & 9e-3\\
         1 & 1 & 12e-3 & 1 & 18e-3\\
   \hline
\end{tabular}\label{tab:instrument_class}
\end{center}
\end{table}

\begin{algorithm}
\caption{Raw-data generation with a realistic noise.}\label{alg:GenData}
\begin{algorithmic}[1]
\Require {Admittance: $\mathbf{Y}$, nodal power injections: $\mathbf{P}, \mathbf{Q}$}
\Procedure{GenData}{}
\For{$k=1:K$}
\State $[\mathbf{V}_{k}, \mathbf{I}_k]$ = LoadFlow($\mathbf{P}_k, \mathbf{Q}_k$, $\mathbf{Y}$)
\State $[\mathbf{\tilde{V}}_k, \mathbf{\tilde{I}}_k]$ =
    \For{$\beta = [\mathbf{V}_k, \mathbf{I}_k]$}
        \State $\delta^m = \mathcal{N}(0, {\sigma^m}|\beta|/3)$
        \State $|\beta| = |\beta| + \delta^m$
        \State $\delta^{p} = \mathcal{N}(0, {\sigma^p}/{3})$
        \State $\text{arg}(\beta) = \text{arg}(\beta) + \delta^{m}$
        \State $\beta = |\beta|\text{exp}(j~\text{arg}(\beta))$
    \EndFor
\State $\mathbf{\tilde{P}}_k+ j\mathbf{\tilde{Q}}_k$ = $\mathbf{\tilde{V}}_k\mathbf{\tilde{I}}_k^*$
\EndFor
\EndProcedure
\end{algorithmic}
\end{algorithm}

\subsubsection{Training, validation, and testing} The 24-hour dataset is split into a training and a validation set {(9 hours each)}. The testing was performed on the last 6 hours of the dataset.
\subsection{Sensitivity Coefficient Estimation Results}
\subsubsection{Estimation Performance Comparison with Different Methods}
\begin{figure}[!htbp]
    \centering
    \subfloat[$K^p_{11,9}$]{\includegraphics[width=\columnwidth]{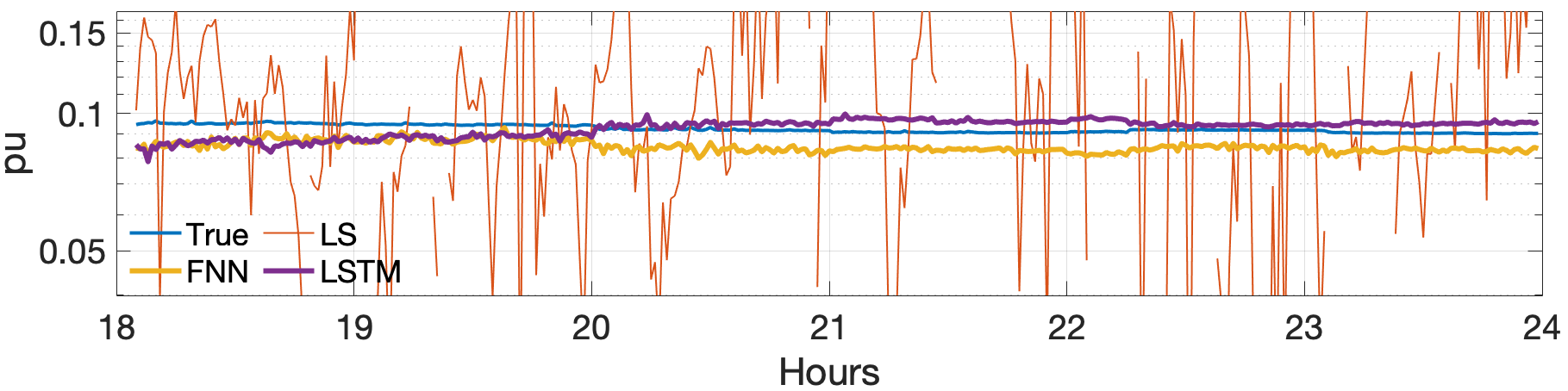}
    \label{fig:Coeff1}}\\
    \subfloat[$K^p_{11,10}$]{\includegraphics[width=\columnwidth]{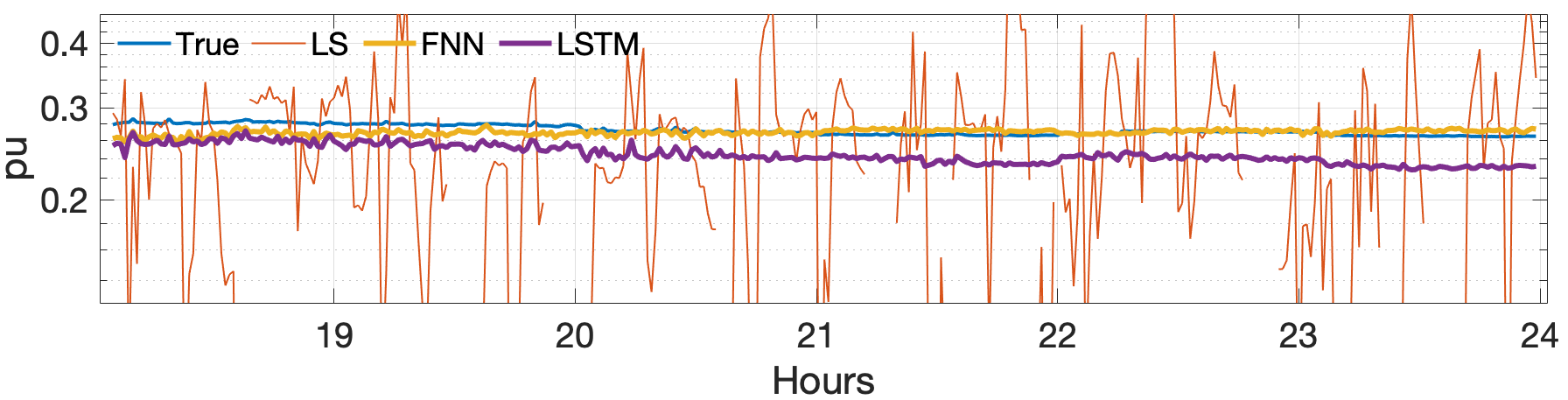}
    \label{fig:Coeff2}}\\
    \subfloat[$K^q_{11,9}$]{\includegraphics[width=\columnwidth]{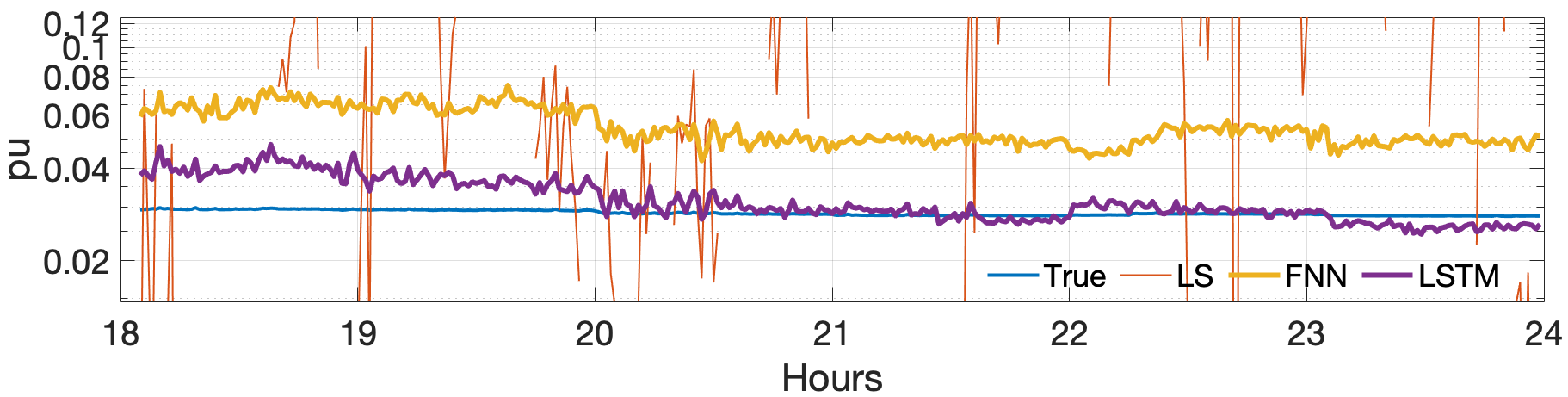}
    \label{fig:coeff3}}\\
    \subfloat[$K^q_{11,10}$]{\includegraphics[width=\columnwidth]{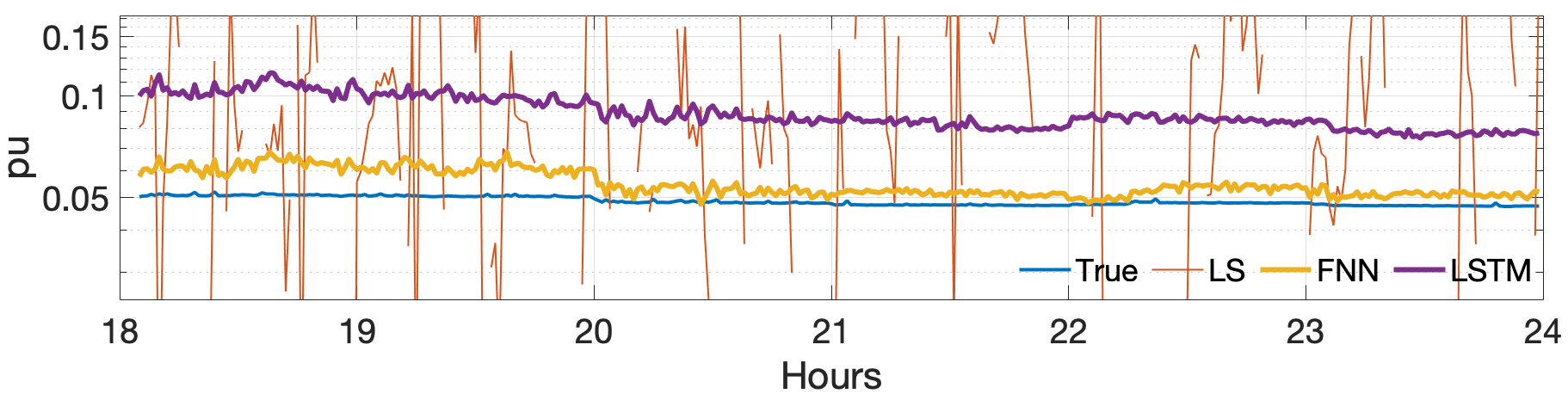}
    \label{fig:coeff4}}
    \caption{Estimated voltage sensitivity coefficients using LS, FNN and LSTM compared against true ones.}
    \label{fig:estimated_SC}
\end{figure}

We present estimation results for sensitivity coefficients in Fig.\ref{fig:estimated_SC}. These results are demonstrated with measurements corresponding to IT class 0.5. For brevity, we showcase the results for a subset of the coefficients. Specifically, Figs.~\ref{fig:Coeff1}, \ref{fig:Coeff2}, \ref{fig:coeff3}, and \ref{fig:coeff4} illustrate the estimation results for $K^p_{11, 9}$, $K^p_{11, 10}$, $K^q_{11, 9}$, and $K^q_{11, 10}$, respectively. The estimations using the LS, FNN, and LSTM are shown in red, orange, and violet colors, respectively, and are compared against the true values shown in blue. It's worth noting that LS estimates exhibit significant variation compared to NN-based schemes; therefore, we have employed a logarithmic y-scale. Log-scale results in missing data from LS as it takes negative values.

\begin{figure*}[!htbp]
    \centering
    \includegraphics[width=0.9\linewidth]{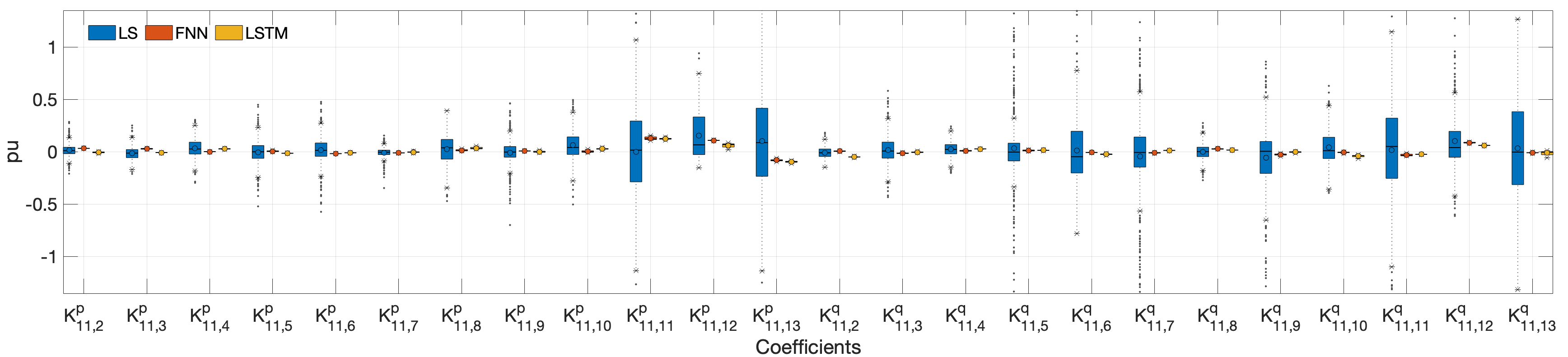}
    \caption{Boxplot of error distributions on different sensitivity coefficients estimated using LS, FNN, and LSTM.}
    \label{fig:boxplot_compare}
\end{figure*}

From the plots, it is evident that both NN-based methods outperform LS for all coefficients. This observation is further supported by the comparison presented in Table~\ref{tab:error_table}, which shows the mean of the root-mean-square-error (mean-of-RMSE) estimates for the node 11. Both the FNN and LSTM-based schemes perform ten times better than LS. Additionally, we illustrate the comparison of error distributions through a boxplot in Fig.~\ref{fig:boxplot_compare}. It is evident from the boxplot that the NN-based methods perform the best.

\begin{table}[!htbp]
    \centering
    \caption{Mean of RMS error on voltage sensitivity coefficients estimates for node 11 (for 6 hours estimates).}
    \begin{tabular}{|c|c|c|c|}
    \hline
        \bf IT class \cite{IT_C, IT_V} & \bf LS (pu) & \bf FNN (pu) & \bf LSTM (pu)  \\
        \hline
        \hline
       0.2 &  0.1656  &  0.0229  &  0.0236\\
       0.5 &    0.2387   &  0.0292  &  0.0304 \\
        1.0 &   0.3264  &  0.0581  &  0.0477 \\
           \hline
    \end{tabular}
    \label{tab:error_table}
\end{table}
\subsubsection{Sensitivity with noise}
We also assess the estimation performance with measurement noise characterized by different IT classes. This is achieved by introducing measurement noises from various IT classes using Algorithm~\ref{alg:GenData}. The performance comparison is presented through the mean of RMSE in Table~\ref{tab:error_table} for different IT classes, highlighting that NN-based methods consistently exhibit lower RMSE.

\begin{figure}[!htbp]
    \centering
    \includegraphics[width=\linewidth]{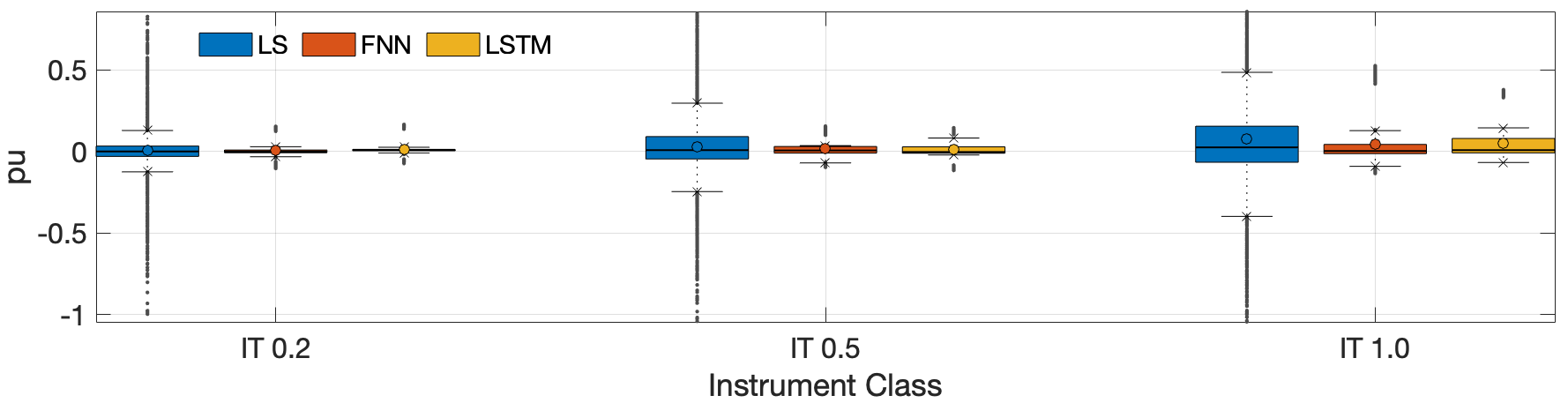}
    \caption{Estimation error comparison with increasing measurement noise for LS, FNN and LSTM.}
    \label{fig:ErrowNoise}
\end{figure}

Furthermore, we provide a visual comparison via boxplots in Fig.~\ref{fig:ErrowNoise} for LS, FNN, and LSTM, respectively. Each boxplot contains the errors corresponding to the coefficients for node $11$. From the plot, it is evident that for LS, the estimation error increases with growing noise, whereas with NN-based methods, the increase in estimation error is minimal with increasing noise. This demonstrates the robustness of the estimation schemes with respect to measurement noise.

\subsubsection{Impact of training dataset length}
\begin{figure}[!htbp]
    \centering
    \includegraphics[width=0.95\linewidth]{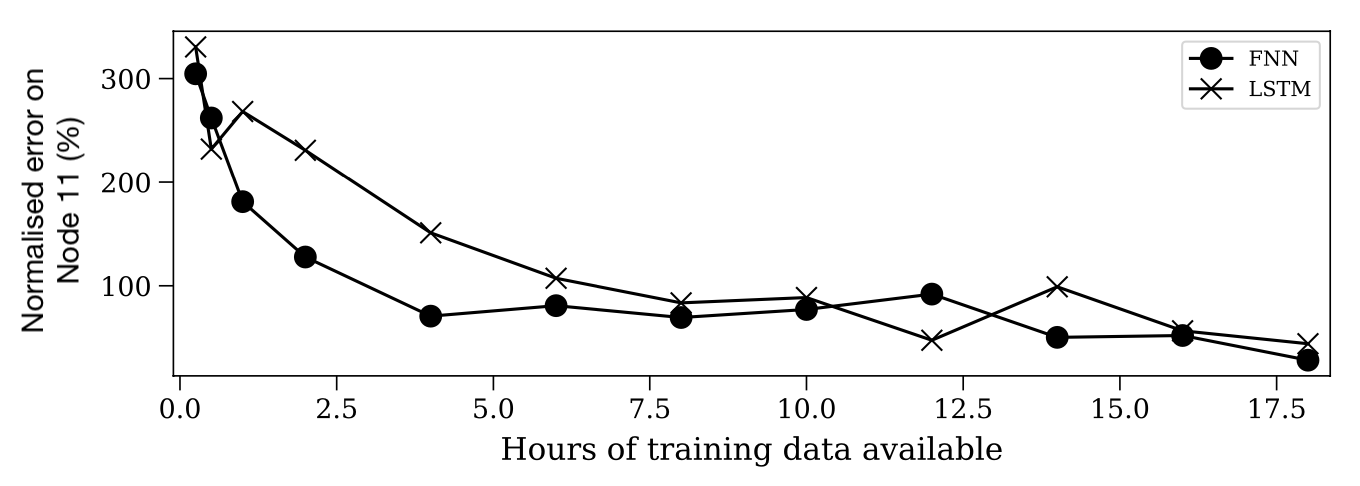}
    \caption{Estimation error with training data length.}
    \label{fig:trainingSize}
\end{figure}

We also explore the impact of training dataset length on estimation performance. To investigate this, we utilized 24 hours of data, reserving the final 6 hours for testing. Subsequently, we trained the NN with different lengths of historical data. The results are depicted in Fig~\ref{fig:trainingSize} for coefficients corresponding to node 11. The x-axis represents the hours of data used for training, with 50\% dedicated to actual training and the remaining 50\% for selecting the best model via validation. The plot illustrates average estimation errors for the coefficient vector corresponding to node 11. The error metric is defined as
    $100 \times \frac{1}{2 (n_b - 1)T} \sum_{t=1}^T\sum_{j = 1}^{2(n_b-1)}\sum_{s\in{p,q}}\Bigg|\frac{K_{ij}^s - \hat{K}^s_{ij}}{{K^s_{ij}}}\Bigg|$.

As observed in Fig~\ref{fig:trainingSize}, the normalized error decreases with an increase in the amount of historical data. Additionally, we observe that estimation errors decrease exponentially with the first hours of training data available. This suggests that in the case of a new topology or network, such neural networks could be trained and made ready for deployment with just 4 hours of 1-second resolution measurements.

\section{Conclusion}
\label{sec:conclusion}
This work introduces a neural network-based approach for estimating voltage sensitivity coefficients, addressing the limitations of poor estimation performance in the presence of increasing measurement noise faced by conventional regression-based methods. Two neural network methods, feedforward, and long-short term memory, are presented and numerically validated on a CIGRE benchmark network.

The numerical results indicate that conventional coefficient estimation methods often encounter numerical stability issues, exhibit high estimation variance, and are highly sensitive to measurement errors. In contrast, both proposed neural network-based estimation methods exhibit almost ten times better accuracy compared to the least squares, demonstrating lower estimation variance and improved numerical stability.

\bibliographystyle{IEEEtran}
\bibliography{Bibliography}
\end{document}